\documentstyle[aps,prl,epsf]{revtex}
\begin{document}
% for two column  activate the line below...
\twocolumn[\hsize\textwidth\columnwidth\hsize\csname@twocolumnfalse\endcsname
\author{G.\ Karapetrov$^a$, M.\ Iavarone$^{a,b}$, W.\ K.\ Kwok$^a$, G.\ W.\ Crabtree$^a$, and D.\ G.\ Hinks$^a$}
\address{$^{a\,}$Materials Science Division, Argonne National Laboratory,
Argonne, Illinois 60439;\\
$^{b\,}$INFM - Dipartimento di Scienze Fisiche of the University
of Naples ``Federico II" , Piazzale Tecchio 80, 80125 Naples,
Italy}
%\author{}
%\address{}
\title{Scanning Tunneling Spectroscopy in MgB$_2$}
\date{\today}
\tighten \maketitle
\begin{abstract}
We present scanning tunneling microscopy measurements of the
surface of superconducting MgB$_2$ with a critical temperature of
39K. In zero magnetic field the conductance spectra can be
analyzed in terms of the standard BCS theory with a smearing
parameter $\Gamma$. The value of the superconducting gap is 5.2
meV at 4.2 K, with no experimentally significant variation across
the surface of the sample. The temperature dependence of the gap
follows the BCS form, fully consistent with phonon-mediated
superconductivity in this novel superconductor. The application of
a magnetic field induces strong pair-breaking as seen in the
conductance spectra in fields up to 6 T.
\end{abstract}

% for two column  activate the line below...
\vskip.2pc]
\narrowtext
%\vspace{-0.4in}
%\begin{multicols}{2}

One route for finding new superconducting compounds with a high
critical temperature combines light elements with high ionicity, a
large density of electronic states at the Fermi level, and stiff
elastic response due to high frequency phonon modes. Many carbides
and nitrides fall into this category, and some of them show
superconducting transition temperatures as high as 10-20 K. The
recent discovery of superconductivity at 40 K in MgB$_2$
\cite{akimitsu} reinvigorates strong interest in this approach.

MgB$_2$ has a remarkably high T$_c$ for a simple binary compound.
The chemical structure is quite simple as well, consisting of
alternating hexagonal layers of Mg atoms and boron honeycomb
layers. Band structure calculations~\cite{bandstructure,band2}
show that this compound is ionic, has a high density of states at
the Fermi level, high phonon frequencies, and strong
electron-phonon interactions. These features favor a high
superconducting transition temperature arising from phonon
mediated electron pairing.  The presence of the isotope
effect\cite{budko} confirms the important role of phonons in the
superconductivity of this compound.  Transport and magnetic
measurements~\cite{canfield,takano,viera,chu,cava} are beginning
to probe the macroscopic response of the superconducting and
normal states, and the supercondicting performance for
applications is being evaluated~\cite{larbalestier,canfieldwires}.

Here we report the superconducting energy gap of polycrystalline
MgB$_2$ pellets as seen in scanning tunneling spectroscopy.  These
measurements directly probe the quasiparticle excitations near the
Fermi energy and provide key information on the nature of the
superconducting energy gap and its temperature and field
dependence. We achieve clean vacuum tunneling with conductance
spectra that are identical within experimental error across the
sample scan area. The tunneling spectroscopy at 4.2 K is
consistent with the modified BCS density of states~\cite{dynes}
(DOS) with a superconducting gap value of 5.2 meV and
pair-breaking strength $\Gamma$ of 3 meV. From STM measurements we
show that the temperature dependence of the gap follows a BCS
behavior. In addition, we present the magnetic field dependence of
the tunneling spectra at 4.2 K showing the pair-breaking effect of
the magnetic field in this material.

The MgB$_2$ sample was synthesized from a high purity, 3 mm
diameter Mg rod and isotopic $^{11}B$ (Eagle Picher, 98.46 atomic
\% $^{11}B$). The Mg rod was cut into pieces about 4 mm long and
mixed with the \-200 mesh $^{11}B$ powder. The reaction was done
under moderate pressure (50 bars) of ultra-high purity argon at
850$^\circ$C. At this temperature the gas-solid reaction was
complete in about one hour. The sample was contained in a machined
BN crucible (Advanced Ceramics Corp. HBC grade BN) with a closely
fitting cover.  There was no reaction between the BN crucible and
the reactants at the synthesis temperature.  X-ray diffraction
showed no impurity peaks in the resultant powder.  The powder was
compacted in a steel die at about 3 Kb and refired using the same
conditions to form the compacted pellet.  The transition
temperature of the material was 39 K measured by DC field cooled
magnetization in an applied field of 2 Oe. Part of this pellet was
used for our STM measurements without additional treatment of the
as-grown surface. The samples were exposed briefly to air before
being inserted in an inert helium atmosphere in the STM. The
tunneling measurements were performed with a home-built STM
operating in helium exchange gas with an electrochemically etched
tip of Pt-Ir wire. The nature of the last atom on the tip was
unknown.

Topographic images over a scale of 300 $\times$ 300 nm$^2$ showed
a flat zone with surface roughness less than 1 nm. To confirm the
vacuum nature of the tunneling we measured the relationship
between the value of the tunneling current and the tip
displacement from the surface. The exponential behavior of this
dependence was verified and indicated a clean vacuum tunneling.
The extracted value of the apparent work
function~\cite{workfunction} is a few hundred meV. The low value
of the tunneling barrier is familiar in the boride family of
compounds although the possibility of modified layer on the
surface of the MgB$_2$ grain cannot be ruled out.

\begin{figure}
\epsfxsize=3.1in \epsffile{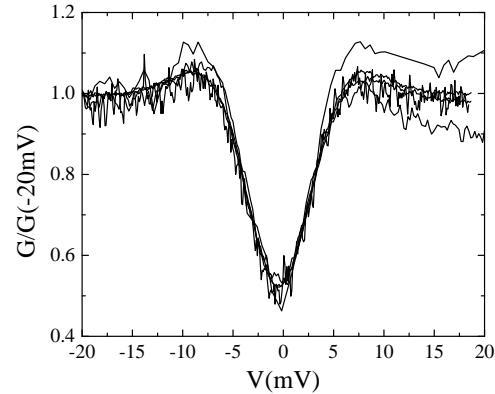} \caption{Conductance spectra
measured as a function of tip-sample spacing  at 4.2 K in zero
magnetic field. The junction resistance values are  0.125, 0.25,
0.5, 1, 2, 3 G$\Omega$. All the spectra have been normalized to
the conductance value at -20 mV.} \label{Figure1}
\end{figure}

Current-voltage $I(V)$ and differential conductance $dI/dV(V)$
curves were recorded using standard lock-in techniques with a
small modulation voltage superimposed on the slowly varying DC
bias voltage while the distance between the tip and sample was
kept constant. The amplitude of the modulation was kept at 0.4 mV,
i.e. three times smaller than the intrinsic thermal broadening
(=3.5k$_B$T) at 4.2 K. Figure~\ref{Figure1} shows a series of
dI/dV(V) curves normalized at V= -20 mV for tunneling resistance
ranging between 125 M$\Omega$ and 2 G$\Omega$ in zero magnetic
field. The tunneling spectra are independent of the junction
resistance, an additional verification of a clean vacuum tunnel
junction. The remarkable reproducibility of the coherence peak and
the value of the zero-bias conductance across the sample surface
is in sharp contrast to STM measurements in cuprate
superconductors. Thermal cycling to 50 K did not change the
junction characteristics.

\begin{figure}
\epsfxsize=3.1in \epsffile{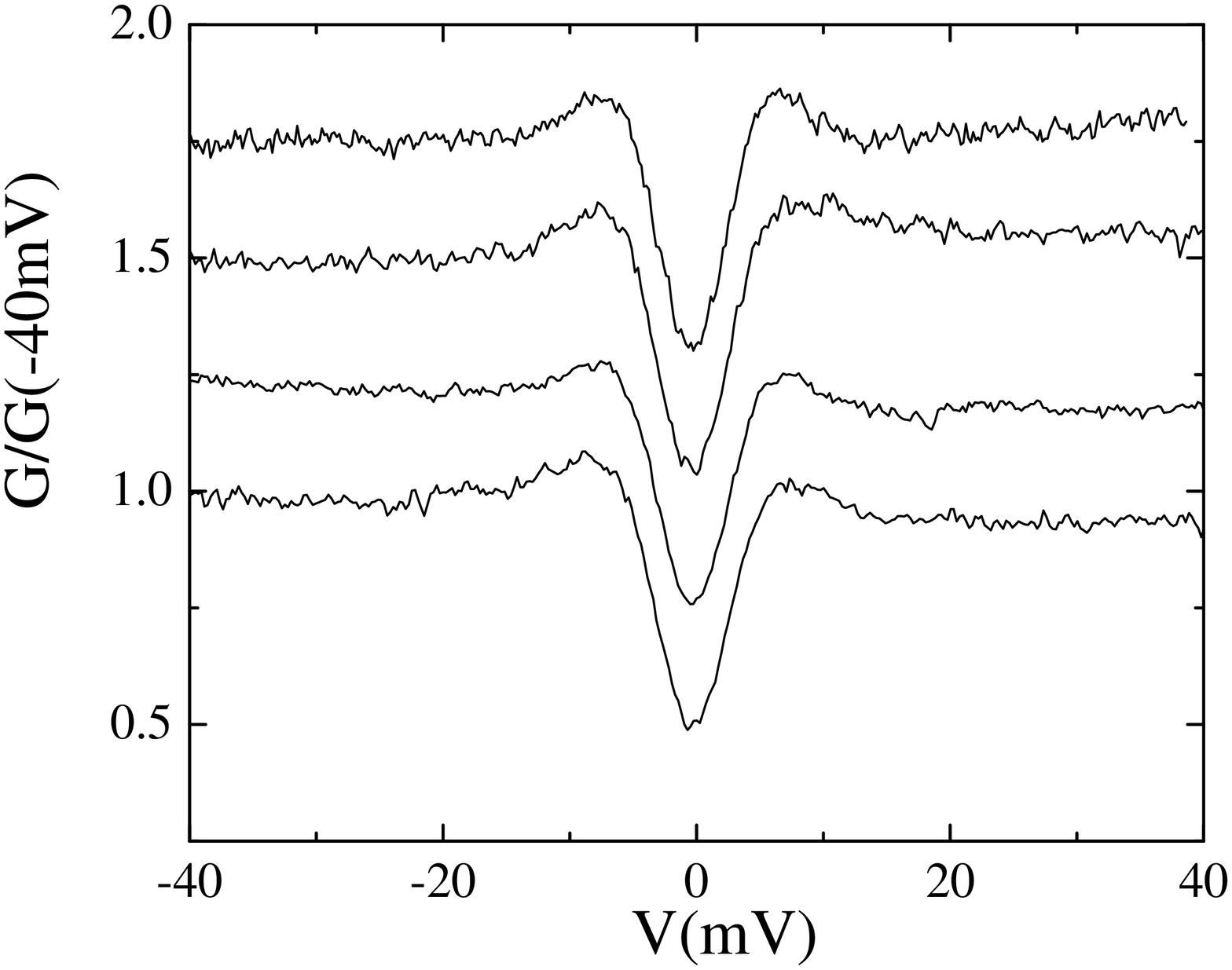} \caption{Tunneling
conductance spectra measured on the scanning area of 300 x 300
nm$^2$ show excellent reproducibility (for clarity the spectra are
equidistantly shifted vertically and normalized to the conductance
value at -20 mV). Junction resistance is 250 M$\Omega$.}
\label{Figure2}
\end{figure}

Unlike the cuprate superconductors, the conductance spectra in
MgB$_2$ are very symmetric implying a BCS-like energy gap. It is
important to note a relatively high zero-bias conductance which is
approximately 50\% of the normal background, and the broad
coherence peak at V=$\pm$ 8 mV. The zero-bias value and the
intensity of the peaks are absolutely reproducible at different
tunneling junction resistances and different locations
(Fig.~\ref{Figure2}) on the sample surface within the scanning
area 300 x 300 nm$^2$.  We exclude RF noise as the cause of such
broadening~\cite{vieraRF} since measurements performed with the
same set-up on Pb and Nb superconducting samples show much lower
zero bias values and much less broadening even though the reduced
temperature T/T$_c$ is much higher. The sizable zero-bias
conductance is accompanied by quite linear conductance
dependence inside the gap region that might suggest the
possibility of coverage of the MgB$_2$ superconductor with a thin
layer of normal material. This case is very similar to the one
observed~\cite{largezbc}  in YNi$_2$B$_2$C and LuNi$_2$B$_2$C in
which the large zero-bias conductance was attributed to tunneling
through an overlayer of normal material.

\begin{figure}
\epsfxsize=3.1in \epsffile{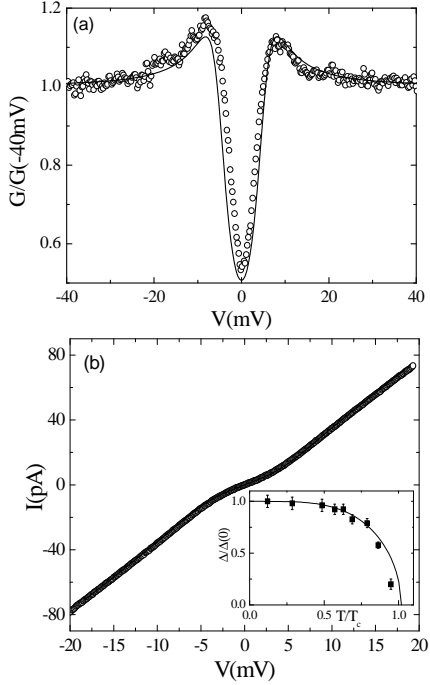} \caption{Scanning tunneling
characteristics on the MgB$_2$ surface at 4.2 K: (a) experimental
tunneling conductance (circles) and calculated conductance curves
using a smeared BCS density of states (solid line) give an
estimate of $\Delta$ = 5.2 meV and $\Gamma$ = 3 meV; (b)
experimental current-voltage characteristic of the same tunneling
junction (circles) and its fit to the same model with identical
parameters (solid line). Insert shows the temperature dependence
of the superconducting energy gap extracted from tunneling
conductance curves at different temperatures (points) following
the conventional BCS behavior (line).} \label{Figure3}
\end{figure}

The value of the gap was estimated using the standard expression
for tunneling conductance~\cite{wolf}  with the modified BCS
expression for the superconducting density of states from the
phenomenological model by Dynes et al.\cite{dynes}:

\begin{equation}
N(E)={\rm Re} \left[ {\frac{ \left|E-\imath \Gamma \right|}{
\sqrt{(E-\imath \Gamma)^2-\Delta ^2}}}\right]
\end{equation}

where $\Delta$ is the superconducting energy gap and $\Gamma$ is
the pair-breaking strength. In Figure~\ref{Figure3}a we show an
example of the experimental tunneling conductance spectrum and
calculated conductance using the above approximation. The best fit
at 4.2K produces $\Delta$=5.2 meV and $\Gamma$=3 meV. The ratio
$2\Delta /kT_c$ = 3 for the bulk value of $T_c$ = 39 K, indicating
a weakly coupled superconductor. Our fits described below suggest
that the transition temperature and the gap in the top layer of
the crystalline surface of the grain are slightly suppresed,
bringing the bulk value of $2\Delta /kT_c$ nearly in line with the
BCS value of 3.5. The value of $\Gamma$=3 meV is consistent with
the electron scattering rate expected for a normal state
resistivity of the order of 1 $\mu \Omega$cm and the band
structure value of the plasma frequency $\Omega_p$=7
meV~\cite{bandstructure}. If so ,the pronounced V-shaped gap we
observe can not be attributed to $d$-wave pairing symmetry,
because the large ratio of $\Gamma / \Delta$ would strongly
suppress T$_c$ for non-s-wave pairing.

Our results show a superconducting gap significantly larger than
the one obtained by Rubio-Bollinger et al.~\cite{viera}  on
isolated grains of commercially prepared material. In addition to
a larger gap, our tunneling spectra show a pronounced V-shape with
many more subgap states than the relatively flat bottomed form of
Rubio-Bollinger et al. These differences cannot be easily
explained by simple surface layer effects. Rather, the
qualitatively different spectra seem to reflect intrinsic
differences in the superconducting nature of the sample surface,
perhaps due to the different preparation histories of the two
samples.

The temperature dependence of the superconducting gap was
extracted from tunneling spectra at a series of temperatures, as
shown in the inset of Fig.~\ref{Figure3}b. The fitting curve and
the experimental points have been normalized to the gap value at
4.2 K ( $\Delta$=5.2 meV) and a T$_c$=35 K. This confirms our
original idea that the origin of the broadening is rather due to a
depressed superconducting layer on the surface. The data are fit
remarkably well with a standard BCS form, further supporting the
importance of phonons for the pairing interaction.

\begin{figure}
\epsfxsize=3.1in \epsffile{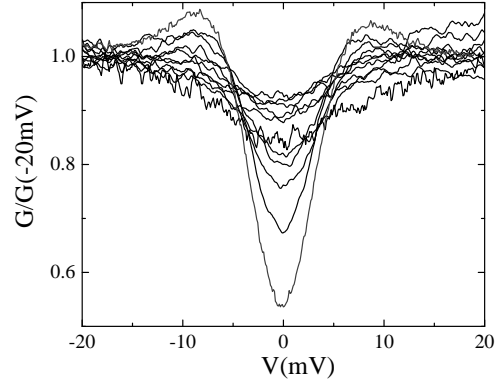} \caption{Magnetic field
dependence of the normalized tunneling conductance at 4.2 K. The
zero bias conductance sequentially increases with the magnetic
field for fields H=0 T, 0.5 T , 0.75 T, 1 T, 2T, 3T, 4T, 5T and
6T. } \label{Figure4}
\end{figure}

It has been shown~\cite{canfield}  that MgB$_2$ is a type II
superconductor and therefore the magnetic flux penetrates the
material in the form of vortices. The superconducting order
parameter inside the vortex core is suppressed, providing
additional quasiparticle states below the zero-field
superconducting gap energy. These states inside the core increase
the zero-bias conductance and smear the coherence peaks in the
conductance spectra. The spectral differences inside the vortex
core enable real space imaging of the vortex configurations in
type II superconductors with STM ~\cite{hess,fisher,yannick}. We
have examined tunneling spectra as a function of location on the
surface in applied magnetic fields and found remarkably
reproducible spectra that gave no indication of an order parameter
variation induced by vortices. This could indicate that natural
pinning in the material is low and the vortices tend to be pinned
in the grain boundaries. Alternatively, if the pinning is weak,
the tunneling current near the tip could disturb the static
position of vortices when scanning, masking their presence.
Although individual vortices were not resolved, the tunneling
conductance spectra taken at different magnetic fields and
different locations show the spatially averaged pair-breaking
effect of the magnetic field (Fig.~\ref{Figure4}). The magnetic
field dramatically increases the number of quasiparticle states in
the gap and smears the superconducting peaks. No additional
features in the gap were observed in applied field.

In conclusion, we performed extensive scanning tunneling
microscopy and spectroscopy measurements on the surface of
superconducting MgB$_2$. Remarkably consistent conductance spectra
were observed across the surface of the sample from which a
superconducting gap of 5.2 meV at 4.2 K was extracted. The
temperature dependence of the gap follows the BCS expression
strongly indicating the importance of phonon-mediated
superconductivity in this material. Applied magnetic fields
produce a strong pair-breaking effect on the conductance spectra
increasing the number of quasiparticle states in the gap.

We would like to thank H.\ Claus and U.\ Welp for T$_c$
measurements and J.\ D.\ Jorgensen, U.\ Welp, M.\ Norman, M.\ Randeria and
A.\ Koshelev for useful discussions. This work was supported by
US DOE Basic Energy Science - Material Science under contract
No. W-31-109-ENG-38. M.I. would like to acknowledge the support of
INFM and the Maria Goeppert-Mayer scholarship at Argonne National
Laboratory.
%\vspace{-.4in}

%\end{multicols}
\end{document}